\documentstyle[12pt,epsfig]{article}

\newcommand{\lsim}{\raisebox{-0.13cm}{~\shortstack{$<$ \\[-0.07cm] $\sim$}}~}

\newcommand{\ra}{\rightarrow}

\newcommand{\s}{\\ \vspace*{-3mm} }
\newcommand{\nn}{\noindent}
\newcommand{\non}{\nonumber}
\newcommand{\beq}{\begin{eqnarray}}
\newcommand{\eeq}{\end{eqnarray}}

\newcommand{\tb}{\tan\beta}
\newcommand{\ctb}{\mbox{ctg}\beta}

\newcommand{\ctg}{\mbox{ctg}}
\newcommand{\ct}[1]{c_{\theta_#1}}
\newcommand{\st}[1]{s_{\theta_#1}}

\newcommand{\czt}[1]{c_{2\theta_#1}}
\newcommand{\szt}[1]{s_{2\theta_#1}}
\newcommand{\tht}[1]{\theta_#1}
\newcommand{\mg}{{m_{\tilde{g}}}}
\newcommand{\tq}{\tilde{q}}
\newcommand{\vivjpaiaj}{(v_{\tq}^i v_{\tq'}^j+a_{\tq}^i a_{\tq'}^j)}
\newcommand{\vivjmaiaj}{(v_{\tq}^i v_{\tq'}^j-a_{\tq}^i a_{\tq'}^j)}
\newcommand{\viajpaivj}{(v_{\tq}^i a_{\tq'}^j+a_{\tq}^i v_{\tq'}^j)}
\newcommand{\viajmaivj}{(v_{\tq}^i a_{\tq'}^j-a_{\tq}^i v_{\tq'}^j)}

\begin{document}

\def\thefootnote{\fnsymbol{footnote}}

\begin{flushright}
PM--96--40\\
KA--TP--30--96\\
hep-ph/9702426\\
\end{flushright}

\vspace{1cm}

\begin{center}

{\large\sc {\bf SUSY Higgs Boson Decays into Scalar Quarks: }}

\vspace*{3mm}

{\large\sc {\bf QCD Corrections}} 

\vspace{1cm}

{\sc A. Arhrib$^{1,2}$, 
A. Djouadi$^{1}$, W. Hollik$^{3}$} and {\sc C. J\"unger$^3$}

\vspace{1cm}

$^1$ Laboratoire de Physique Math\'ematique et Th\'eorique, UPRES--A 5032,\\
Universit\'e de Montpellier II, F--34095 Montpellier Cedex 5, France.

\vspace*{3mm}

$^2$ Laboratoire de Physique et Techniques  Nucl\'eaires  \\ 
Facult\'e des Sciences Semlalia, B. P S15, Marrakech Morocco.

\vspace*{3mm}

$^3$ Institut f\"ur Theoretische Physik, Universit\"at Karlsruhe,\\
D--76128 Karlsruhe, Germany. 

\end{center}

\vspace{2cm}

\begin{abstract}

\nn In supersymmetric theories, the decays of the neutral CP--even and CP--odd 
as well as the charged Higgs bosons into scalar quarks, in particular into top 
and bottom squarks, can be dominant if they are kinematically allowed. 
We calculate the QCD corrections to these decay modes in the minimal 
supersymmetric extension of the Standard Model, including all quark mass 
terms and squark mixing. These corrections turn out to be rather large, 
altering the decay widths by an amount which can be larger than 50\%. 
The corrections can be either positive or negative, and depend strongly
on the mass of the gluino. We also discuss the QCD corrections
to the decays of heavy scalar quarks into light scalar quarks and Higgs bosons. 
 
\end{abstract}

\newpage

\def\thefootnote{\arabic{footnote}}
\setcounter{footnote}{0}

\subsection*{1. Introduction}

Supersymmetric theories \cite{R1} are widely considered as the most 
attractive extensions of the Standard Model. In a grand unified
framework, they protect scalar Higgs bosons from acquiring very large 
masses and provide the opportunity to generate the electroweak symmetry
breaking radiatively. In the minimal version of 
these theories, the Minimal Supersymmetric Standard Model (MSSM), the 
Higgs sector is extended to contain two doublet fields, leading to five 
physical states \cite{R2}: two CP--even neutral Higgs bosons, $h$ and $H$, 
a CP--odd Higgs boson $A$ and two charged Higgs particles $H^\pm$. While
the lightest CP--even Higgs boson $h$ mass is predicted to be less than 
$130$ GeV \cite{R3}, the $H,A$ and $H^\pm$ states are expected to have 
masses of order of the electroweak symmetry breaking scale. \s

If all genuine SUSY particles are very heavy, the neutral and charged 
Higgs bosons will decay into standard fermions and gauge bosons, as well as 
into cascades involving the lighter Higgs bosons; in most cases the decays 
into heavy $b$ and $t$ quark pairs are dominant. These decay modes have 
been extensively discussed in the literature and for a recent summary we 
refer the reader to Ref.~\cite{R4}. However, it could well be that charginos, 
neutralinos and sfermions are light enough for the Higgs decays into these 
particles to be kinematically allowed. For instance, in  grand unified 
models with proper radiative electroweak symmetry breaking, the $H,A$
and $H^\pm$ bosons are rather heavy and approximately mass degenerate,
$M_H \simeq M_A \simeq M_{H^\pm} \sim$ a few hundred GeV, while charginos, 
neutralinos and eventually top and bottom squarks have masses of ${\cal 
O}$(100 GeV) \cite{R5}. In this case, the decay pattern of the heavy Higgs 
particles $H,A$ and $H^\pm$ will be drastically different \cite{R6a,R6b}. In 
particular, because of the large Yukawa couplings and large squark mixing, 
the decays of the heavy Higgs bosons into top and bottom squarks can be 
competitive with the standard decay channels and can even be the dominant 
ones in some areas of the MSSM parameter space \cite{R6b}. \s
   
It is well known that the standard hadronic decay modes of Higgs particles
are significantly affected by QCD radiative corrections \cite{R7}. In order 
to have full control on the Higgs decay widths into squark pairs -- and
to make a reliable comparison with the standard decays -- QCD corrections 
have also to be included. Recently, the QCD corrections to the decay of the 
charged Higgs boson into $\tilde{t} \tilde{b}$ pairs have been calculated 
\cite{R8} and found to be quite substantial. \s

In the present paper, we derive the QCD corrections to the decays of the 
neutral CP--even and CP--odd
Higgs bosons into stop and sbottom pairs. These corrections are found to be 
rather large, altering the Born decay widths by an amount which can exceed
50\%. These effects must therefore be taken into account. We also rederive 
the QCD corrections to the decays of the charged Higgs boson into $\tilde{t} 
\tilde{b}$ pairs using a different [and more convenient] renormalization 
scheme compared to the one adopted in Ref.~\cite{R8}. We find that the 
corrections can be extremely large and can strongly suppress the decay
width compared to its tree--level value. Finally, we also discuss for 
completeness the QCD corrections to the reverse decay of heavy squarks 
into their lighter partners plus light Higgs bosons; these decays have been
discussed at tree--level in Ref.~\cite{R10a}. \s

The paper is organized as follows: to set the notation and to introduce the
various parameters which enter the analysis, we summarize in the next section
the Higgs decay widths into squark pairs in Born approximation. In section 
3, we present  the analytic expressions of the QCD corrections to the decay
widths. Section 4 contains the discussion of the numerical results. In section
5, we discuss the QCD corrections to the decays of heavy squarks into 
Higgs bosons and we give our conclusions in section 6.

\subsection*{2. Born Approximation}

The amplitudes for the decay widths of the MSSM neutral heavy CP--even, 
CP--odd and charged Higgs bosons\footnote{In view of the experimental 
bounds on the squark masses \cite{R9}, the lightest CP--even Higgs boson 
in the MSSM cannot decay into squark pairs since its maximal mass value is 
smaller than $M_h \lsim 130$ GeV \cite{R3}. These decays will therefore not 
be considered here. 
However, the analytical expressions can be straightforwardly obtained from 
the decays of the heavy CP--even Higgs boson, after the proper changes in
the Higgs boson couplings.}, that we will collectively denote by $\Phi$, 
into the scalar partners of first and second generation quarks
\beq
\label{phidec}
\Phi \ \ra \ \tilde{q}_i \, \overline{\tilde{q}}'_j
\ \ \ , \ \ \Phi =H,A, H^\pm
\eeq
depend on three parameters if the quark mass is neglected: the mass
of the decaying Higgs boson, the squark masses [more precisely the left-- 
and right--handed soft SUSY breaking scalar masses $m_{\tilde{q}_L}$ and 
$m_{\tilde{q}_R}$, which in general are taken to be equal] and the ratio of 
the vacuum expectation values of the two MSSM Higgs fields $\tb$. A mixing 
angle $\alpha$ in the CP--even Higgs sector also enters the amplitudes, but 
in the MSSM it can be expressed in terms of $\tb$ and the Higgs boson mass
$M_\Phi$. \s

In the case of the third generation scalar quarks, and in particular top 
squarks, the mixing between left-- and right--handed squarks, which is 
proportional to the mass of the partner quark, must be included \cite{R10}.  
In terms of the scalar masses $m_{\tilde{q}_L}, m_{\tilde{q}_R}$, the 
Higgs--higgsino mass parameter $\mu$ and the soft SUSY--breaking trilinear 
coupling $A_q$, the squark mass matrices read
\beq 
\label{sqmass_matrix}
{\cal M}^2_{\tilde{q}} =
\left(
  \begin{array}{cc} m_q^2 + M_{LL}^2 & m_q \,M_{LR} \\
                    m_q\,M_{LR}      & m_q^2 + M_{RR}^2 
  \end{array} \right) \non 
\eeq
\beq
M_{LL}^2 &=&m_{\tilde{q}_L}^2 + (I_3^q - e_q s_W^2)\cos 2\beta\,M_Z^2 \non \\
M_{RR}^2 &=& m_{\tilde{q}_R}^2 + e_q s_W^2\,\cos 2\beta\,M_Z^2 \non \\
M_{LR} &=& A_q+\mu(\tb)^{-2 I_3^q} \ . 
\eeq
$I_3^q$ and $e_q$ are the weak isospin and electric charge of the squark 
$\tilde{q}$, and $s_W^2=1-c_W^2 \equiv \sin^2\theta_W$. They are diagonalized 
by the $ 2 \times 2$ rotation matrices 
\beq
 R^{\tilde{q}} &=&  \left( \begin{array}{cc}
     \ct{q} & -\st{q} \\ \st{q} & \ct{q}
  \end{array} \right)  \ \ \ \ , \ \ \ct{q} \equiv \cos \theta_q 
\ \ {\rm and} \ \ \st{q} \equiv \sin \theta_q 
\eeq
which turn the mass eigenstates, $\tilde{q}_1$ and $\tilde{q}_2$, into
the current eigenstates $\tilde{q}_L$ and $\tilde{q}_R$; the mixing angle and 
squark masses are then given by 
\beq
\tan \theta_q & = &\frac{2\,m_q\,M_{LR}}{ 
      M_{LL}^2 - M_{RR}^2 - \sqrt{(M_{LL}^2-M_{RR}^2)^2 +4\,m_q^2\,M_{LR}^2}}
\eeq
and
\beq
m_{\tilde{q}_{1,2}}^2 = m_q^2 +\frac{1}{2} \left[
M_{LL}^2 +M_{RR}^2 \mp \sqrt{ (M_{LL}^2
-M_{RR}^2 )^2 +4 m_q^2 M_{LR}^2 } \ \right]
\eeq

In the Born approximation, the decay widths of the MSSM heavy 
CP--even, CP--odd and charged Higgs bosons into squark pairs, eq.~(\ref{phidec}),
can be written as \cite{R11}
\beq
\label{phiborn}
\Gamma^0 (\Phi \ra \tilde{q}_i \overline{\tilde{q}}'_j) = \frac{3 G_F }{4
\sqrt{2} \pi \, M_{\Phi}^3} \, \lambda^{1/2} (M_{\Phi}^2,m_{\tilde{q}_i}^2,
m_{\tilde{q}'_j}^2) \; \left( \; G_{\Phi i j} \; \right)^2
\eeq
where $\lambda = x^2+y^2+z^2-2 \, (xy+xz+yz)$ is the two--body phase space
function. The couplings of the Higgs bosons to squarks $G_{\Phi ij}$ read
\beq
\label{couplingmatrix}
G_{\Phi ij } = \frac{1}{\sqrt{2}} \, \sum_{k,l=1}^2 \  
\left( R^{\tilde{q}} \right)_{ik}^{\rm T} \, C_{\Phi \tilde{q}
\tilde{q}' }^{kl} \, \left( R^{\tilde{q}'} \right)_{lj} 
\eeq
with the matrices $C_{\Phi \tilde{q} \tilde{q}' }$ summarizing the couplings 
of the Higgs bosons to the squark current eigenstates; for the $H,A$ and 
$H^\pm$ particles, they are given by 
\beq
\label{h0couplings}
C_{H \tilde{q} \tilde{q}} = \left( \begin{array}{cc}
\left(2I_3^q - 2e_q s_W^2\right) M_Z^2 \cos(\beta+\alpha) + 2 m_q^2 r_1^q & 
m_q (A_q r_1^q + \mu r_2^q) \\ m_q (A_q r_1^q + \mu r_2^q) & 
2 e_q s_W^2 M_Z^2 \cos(\beta+\alpha) + 2 m_q^2 r_1^q 
\end{array} \right) 
\eeq
\beq
\label{a0couplings}
C_{A \tilde{q} \tilde{q}} & = & \left( \begin{array}{cc}
0 & -m_q \left[ \mu - A_q (\tb)^{-2I_3^q} \right] \\
m_q \left[ \mu - A_q (\tb)^{-2I_3^q} \right] & 0 
\end{array} \right) 
\eeq
\beq
C_{H^\pm \tilde{t} \tilde{b}} &= & \sqrt{2} \, \left( \begin{array}{cc}
m_b^2 \tb + m_t^2/\tb - M_W^2 \, \sin 2\beta & m_b \,(A_b \tb  -\mu) \\
m_t \,(A_t/\tb  -\mu) & 2  \,m_t \,m_b/ \sin2\beta
             \end{array} \right)
\eeq
with the coefficients $r^q_{1,2}$ as 
\beq
r_1^t = \frac{ \sin \alpha}{\sin \beta} \ \  , \ \ 
r_2^t = \frac{ \cos \alpha}{\sin \beta} \ \ , \ \
r_1^b = \frac{ \cos \alpha}{\cos \beta} \ \ , \ \ 
r_2^b = \frac{ \sin \alpha}{\cos \beta}\;.
\eeq

\smallskip

In principle, for the Higgs boson masses and the mixing angle $\alpha$,
one has also to include the large radiative corrections \cite{R3} which 
grow as the fourth power of $m_t$ and induce an additional dependence on 
$\mu$ and $A_q$ at the subleading level. However for rather heavy Higgs bosons, 
$M_H \sim M_A \sim M_{H^\pm} \gg M_Z$, these corrections will not affect the 
decay amplitudes: the decaying Higgs boson mass can be used as input parameter
and the angle $\alpha$ will reduce anyway to the value $\alpha \ra \beta-\pi/2$.
In this case, the couplings of the CP--even Higgs boson will simplify to 
\beq
C_{H \tilde{q} \tilde{q}} = 
\left( \begin{array}{cc}
\left(2I_3^q - 2e_q s_W^2\right) M_Z^2 \sin 2\beta \mp 2 m_q^2 (\tb)^{\mp 1} & 
\mp m_q [A_q  (\tb)^{\mp 1}  - \mu ] \\ 
\mp m_q [A_q  (\tb)^{\mp 1}  - \mu ] & 
2 e_q s_W^2 M_Z^2 \sin 2\beta \mp 2 m_q^2 (\tb)^{\mp 1} \non 
\end{array} \right) 
\eeq
where the ``--'' is for up--type squarks and the ``+'' for down--type squarks.  
When the quark masses and the squark mixing angles are set to zero, as is
the case for first and second generation squarks, the pseudoscalar Higgs
boson cannot decay at tree-level into squark pairs since the $A \tilde{q}_i 
\tilde{q}_i$ coupling is zero by virtue of CP--invariance and the $A 
\tilde{q}_1 \tilde{q}_2$ coupling is proportional to $m_q$. \s

The decay widths of the heavy neutral 
CP--even and the charged Higgs bosons into first and second generation 
squarks are proportional to 
$$
\Gamma^0 \sim  G_F M_W^4 \sin^2 2\beta 
/M_{\Phi}
$$ 
in the asymptotic regime $M_\Phi \gg m_{\tilde{q}}$. These decays are 
suppressed by the heavy Higgs mass and therefore cannot compete with the
dominant decay modes into top and/or bottom quarks [and to charginos and 
neutralinos] for which the decay widths grow as $M_\Phi$. \s
 
In contrast, for the decays involving top squarks, the partial widths 
up to mixing angle factors are proportional  to 
$$
\Gamma^0 \sim G_F m_t^4/ (M_\Phi \tan^2\beta) \ \ {\rm or/and} 
\ \ G_F m_t^2 (\mu+ A_t/\tb)^2 /M_\Phi \;.
$$
For small $\tb$ values and  not too heavy Higgs bosons, or for intermediate
values of $\tb$  and for $\mu$ and $A_t$ values of the order of $\sim 1$ TeV, 
the partial decay widths into top squarks can be very large and can compete 
with, and even dominate over, the decay channels into top quarks [and into
charginos/neutralinos]. Furthermore, decays into sbottom quarks can also be 
important for large values of $\tb$ and $A_b$. We therefore focus in this 
paper on the Higgs boson decays into third generation squarks; however, since we
give complete analytic expressions, the decay widths into first and second
generation can straightforwardly be obtained by setting the quark masses
and the squark mixing angles to zero. 

\subsection*{3. QCD Corrections} 

The QCD corrections to the Higgs decays into scalar quarks, eq.~(\ref{phidec}),
consist of virtual corrections, Figs.1a--d, and real corrections with an additional 
gluon emitted off the final squark states, Fig.~1e. The ${\cal O}(\alpha_s)$ 
virtual contributions can be split into contributions with gluon [1a] and 
gluino [1b] exchange in the $\Phi \tilde{q} \tilde{q}'$ vertex, the mixing 
contribution due to the off--diagonal self energies of the outgoing squarks 
[1d] and the quartic squark interaction contribution [1c]. These 
contributions have to be supplemented by the wave-function counterterms, and 
by a counterterm renormalizing the $\Phi \tilde{q}_i\tilde{q}'_j$ interaction. 
\s

The partial widths of the decays of the MSSM Higgs bosons into scalar quarks,
eq.~(\ref{phidec}), at ${\cal O}(\alpha \alpha_s)$ can be written as 
\beq
\Gamma^1 (\Phi \ra \tilde{q}_i \overline{\tilde{q}}'_j) =
\frac{G_F} {2 \sqrt{2} \pi \, M_{\Phi}^3} \ \frac{\alpha_s}{\pi} \
     \lambda^{1/2} (M_{\Phi}^2,m_{\tilde{q}_i}^2,m_{\tilde{q}'_j}^2) \; 
      \; G_{\Phi ij} \; \Delta_{\Phi ij}
\eeq
with 
\beq
\label{virtcorr}
\Delta_{\Phi ij}=\Delta^V_{\Phi ij}+\Delta^{\rm CT}_{\Phi ij}
+\Delta^R_{\Phi ij}\;.
\eeq

The the sum of the virtual and counterterm corrections,
$\Delta^V_{\Phi ij}$ and $\Delta^{\rm CT}_{\Phi ij}$, is ultraviolet finite,
as it should be, but still infrared divergent. The calculation has been
performed in the dimensional reduction scheme \cite{R12}, which preserves
supersymmetry at the one loop level. However, the results are the same in both
the dimensional regularization \cite{R13} and dimensional reduction schemes
if the quark mass counterterms, which are needed to renormalize the
$\Phi \tilde{q} \tilde{q}'$ interaction, are shifted appropriately  by a finite
amount \cite{R14}. The infrared divergence, which is regulated by introducing
a fictitious mass $\lambda$ for the gluon, is cancelled after including
the real corrections  $\Delta^R_{\Phi ij}$. 

\subsubsection*{3.1 Virtual Corrections} 

The ${\cal O}(\alpha_s)$ virtual corrections in eq.~(\ref{virtcorr}) can be decomposed
in the following way, 
\beq
\Delta^V_{\Phi ij} = \Delta^{g}_{\Phi ij} + \Delta^{\rm mix}_{\Phi ij} + 
\Delta^{4\tilde{q}}_{ \Phi ij} + \Delta^{\tilde{g}}_{ \Phi ij} 
\eeq
with $\Delta^{g}$,  $\Delta^{\rm mix }$, $\Delta^{4\tilde{q}}$ and
$\Delta^{\tilde{g}}$ the contributions from the diagrams with gluon 
exchange, squark mixing, the quartic squark interaction and the gluino 
exchange respectively. 

\smallskip

The contribution from the \underline{gluon exchange} in the $\Phi \tilde{q}
\tilde{q}'$ vertex is given by
\beq
\Delta^{g}_{ \Phi ij} &=& G_{ \Phi ij} \Big[
   B_0(m_{\tilde{q}_i}^2, \lambda, m_{\tilde{q}_i})
         + B_0(m_{\tilde{q}'_j}^2, \lambda, m_{\tilde{q}'_j})
         - B_0(M_{\Phi}^2, m_{\tilde{q}_i}, m_{\tilde{q}'_j}) \non\\
&&\hspace{1.3cm}
      + 2 \,( m_{\tilde{q}_i}^2 + m_{\tilde{q}'_j}^2 - M_{\Phi}^2 )
          \,C_0(m_{\tilde{q}_i}^2,M_{\Phi}^2,m_{\tilde{q}'_j}^2;
              \lambda,m_{\tilde{q}_i},m_{\tilde{q}'_j})  \Big] 
\eeq
with $B_0$ and $C_0$ the Passarino--Veltman scalar two-- and three point
functions], the expressions of which can be found in Ref.~\cite{R15}. 

\smallskip


The contribution from the \underline{off--diagonal self energies} of the 
external scalar quarks  evaluates to
\beq
\Delta^{\rm mix}_{ \Phi ij} &=&   \frac{G_{\Phi ij'} }
{ m_{\tilde{q}'_{j'}}^2  -m_{\tilde{q}'_j}^2 } 
\Big[ \czt{{q'}} \szt{{q'}} \;
           ( A_0(m_{\tilde{q}'_2}^2)-A_0(m_{\tilde{q}'_1}^2))   
            + 4 \czt{{q'}} m_{q'}  \mg
             B_0(m_{\tilde{q}'_j}^2,\mg,m_{q'})  \Big]  \non\\
&+& \frac{ G_{\Phi i'j } }{ m_{\tilde{q}_{i'}}^2 
- m_{\tilde{q}_i}^2}
     \Big[ \czt{q} \szt{q}
           ( A_0(m_{\tilde{q}_2}^2)-A_0(m_{\tilde{q}_1}^2))  
           + 4 \czt{q} m_q  \mg 
            B_0(m_{\tilde{q}_i}^2,\mg,m_q) \Big] 
\eeq
where  $j' = 3-j$ and $i'=3-i$ and $A_0$ the scalar one--point function
\cite{R15}.  

\smallskip

The contribution from the loop diagram involving the \underline{quartic 
squark interaction} is given by
\beq
\Delta^{4\tilde{q}}_{ \Phi ij} = - \sum_{k,l=1,2}\;
          S^{\tilde{q}}_{ik} \; G_{\Phi k l }\; S^{\tilde{q}'}_{lj}\;
          B_0(M_{\Phi}^2, m_{\tilde{q}_k},m_{\tilde{q}'_l})
\eeq
with the matrix $S^{\tilde{q}}$ defined as
\beq
 S^{\tilde{q}} &=& 
  \left( \begin{array}{cc}
     \czt{q} & -\szt{q} \\ -\szt{q} & -\czt{q}
  \end{array} \right) \;  .
\eeq

\smallskip


Finally, the \underline{gluino exchange} contribution in the $\Phi \tilde{q}
\tilde{q}'$ vertex reads
\beq
\label{gluinovertex}
\Delta^{\tilde{g}}_{\Phi ij} &=& 
\Big\{  v_s (m_q \vivjpaiaj + \mg \vivjmaiaj) \non\\
&&\hspace{.7cm}
              - a_s (m_q \viajpaivj + \mg \viajmaivj)
        \Big\}\,B_0(m_{\tilde{q}_i}^2, \mg, m_q) \non\\
&&\hspace{0.2cm}
 + \;   \Big\{  v_s (m_{q'} \vivjpaiaj + \mg \vivjmaiaj) \non\\
&&\hspace{.7cm}
              + a_s (m_{q'} \viajpaivj - \mg \viajmaivj)
        \Big\}\,B_0(m_{\tilde{q}'_j}^2, \mg, m_{q'}) \non\\
&&\hspace{0.2cm}
  + \;  \Big\{  v_s (m_q \vivjpaiaj + m_{q'} \vivjpaiaj) \non\\
&&\hspace{.75cm}
              - a_s (m_q \viajpaivj - m_{q'} \viajpaivj)
        \Big\}\,B_0(M_{\Phi}^2, m_q, m_{q'}) \non\\
&&\hspace{0.cm}
 + \;   \Big\{
           a_s \, \mg \, \viajmaivj ( M_{\Phi}^2
           -(m_q-m_{q'})^2 )
\non\\ &&\hspace{0.2cm}
       -   v_s \, \mg \, \vivjmaiaj ( M_{\Phi}^2
           -(m_q+m_{q'})^2 )
\non\\ &&\hspace{0.2cm}
       +   a_s \viajpaivj (
            m_{\tilde{q}'_j}^2 m_q
           -m_{\tilde{q}_i}^2 m_{q'}
           - (\mg^2
             - m_q m_{q'} ) (m_q-m_{q'}) )
\non\\ &&\hspace{0.2cm}
       -   v_s \vivjpaiaj (
            m_{\tilde{q}'_j}^2 m_q
           +m_{\tilde{q}_i}^2 m_{q'}
           - (\mg^2
             + m_q m_{q'} ) (m_q+m_{q'}) )
        \Big\}\,
\non\\ &&\hspace{2.2cm} \times
        C_0(m_{\tilde{q}_i}^2,M_{\Phi}^2,m_{\tilde{q}'_j}^2,\mg,m_q,m_{q'}) 
\eeq
Here, $v_s$ and $a_s$ denote the scalar and pseudoscalar couplings of
the Higgs boson $\Phi$ to the quarks, which for $H, A$ and $H^\pm$
are given by
\beq
&H q \bar{q} :& v_s  =  2 \sqrt{2} m_q r_1^q \ \ \ \ \ \ , \ \ a_s=0 \non \\
&A q \bar{q} :& a_s  =  2 \sqrt{2} m_q (\tb)^{-2I_3^q}  \ \ , \ \ v_s=0 \non \\
&H^+ t \bar{b} :& 
v_s/a_s = 2 (m_b\,\tb \pm m_t\,\ctb) 
\eeq
and $v_{\tilde{q}}^i,a_{\tilde{q}}^i$  are the reduced gluino--quark--squark
couplings which read
\beq
v_{\tilde{q}}^1 =a_{\tilde{q}}^2 = \frac{1}{2} (\ct{q}-\st{q}) \ \ 
{\rm and} \ \ 
a_{\tilde{q}}^1 = - v_{\tilde{q}}^2 = \frac{1}{2} (\ct{q}+\st{q}) \;.
\eeq
\smallskip

We have verified that these expressions agree with those given in Ref.~\cite{R8}
in the case of the charged Higgs boson decays into $\tilde{t} \tilde{b}$
final states.

\subsubsection*{3.2 Counterterm corrections} 

The counterterm corrections consist of the external squark wave--function 
renormalization and the renormalization of the Higgs--squark interaction. 
As discussed previously, this interaction involves the quark masses, the 
squark mixing angles and the trilinear squark couplings, and a counterterm 
for each of these parameters will be needed. However, the trilinear coupling 
can be expressed in terms of the quark and squark masses and the squark 
mixing angle  
\beq
A_q  =  \frac{1}{m_q} \, \st{q} \ct{q} \left( m_{\tilde{q}_1}^2 
- m_{\tilde{q}_2}^2  \right) -   \mu ( \tb )^{-2I_3^q} 
\eeq
with the parameters $\mu$ and $\beta$ not renormalized by strong interactions. 
Therefore, the counterterm for the trilinear coupling $\delta A_q$ is fixed 
by the quark and squark mass counterterms $\delta m_q$ and $\delta m_{\tilde{q
}_i}$, and by the mixing angle counterterm $\delta \theta_q$
\beq
\label{AqCT}
 \delta A_q = \frac{ m_{\tilde{q}_1}^2-m_{\tilde{q}_2}^2 }{2m_q}
  \bigg( 2 \czt{q} \, \delta \tht{q}
  - \szt{q} \,\frac{\delta m_q}{m_q} \bigg)
  + \szt{q} \, m_{\tilde{q}_1} \, \frac{\delta m_{\tilde{q}_1} }{m_q}
  - \szt{q} \, m_{\tilde{q}_2} \, \frac{\delta m_{\tilde{q}_2} }{m_q} \; .
\eeq

We perform the renormalization programme in the on--shell scheme where 
the quark and squark masses are defined as the poles of their respective
propagators $\delta m_{\tilde{q}_i}=\Sigma_{ii}^{\tilde{q}} (m_{\tilde{q}}^2)$ 
and $\delta m_q =\Sigma_{ii}^{q}(m_q^2)$. The squark wave-function 
renormalization constants $\delta Z_{\tilde{q}_i}$ are defined as usual 
in such a way that the residues at the poles are equal to one.  \s


Finally, as done in Ref.~\cite{R16a}, the mixing angle counterterms are 
fixed by the requirement that the renormalized self energy for one of the 
squarks should remain diagonal, and we will choose this squark to be the 
lightest one. This is similar to the treatment made in Ref.~\cite{R16b}.

\smallskip

The various counterterms then read 

\beq
\label{d_Zsq}
 \frac{1}{2} \; \delta Z_{\tilde{q}_i} & = &  (m_q^2+\mg^2-m_{\tilde{q}_i}^2) 
\;{B_0}'(m_{\tilde{q}_i}^2,m_q,\mg) - B_0(m_{\tilde{q}_i}^2,m_q,\mg)  
+ B_0(m_{\tilde{q}_i}^2, \lambda ,m_{\tilde{q}_i})
\non\\
&&
   + 2(-1)^i \szt{q}\; \mg m_q\; {B_0}'(m_{\tilde{q}_i}^2,m_q,\mg)
   + 2 m_{\tilde{q}_i}^2\; {B_0}'(m_{\tilde{q}_i}^2, \lambda,m_{\tilde{q}_i}) 
\eeq
\beq
\label{sqm_ct}
m_{\tilde{q}_i} \delta m_{\tilde{q}_i} & = & - (m_q^2 +\mg^2-m_{\tilde{q}_i}^2)
\; B_0(m_{\tilde{q}_i}^2,m_q,\mg) - 2 m_{\tilde{q}_i}^2\; B_0(m_{\tilde{q}_i}^2
, \lambda ,m_{\tilde{q}_i})  \non \\
&& -A_0(\mg) - A_0(m_q) +\frac{1}{2} \left[ (1+  \czt{q}^2 )\;
A_0(m_{\tilde{q}_i}) + \szt{q}^2 \; A_0(m_{\tilde{q}_{i'}}) \right] \non \\
&& - 2(-1)^i \szt{q} \;\mg \,m_q \;B_0(m_{\tilde{q}_i}^2,m_q,\mg) 
\eeq
\beq
\frac{ \delta m_q}{m_q} & = & -\Big[ 2 \, B_1(m_q^2,m_q,\lambda) + 
4 \, B_0(m_q^2,m_q, \lambda)  + B_1(m_q^2,\mg,m_{\tilde{q}_1}) + 
B_1(m_q^2,\mg,m_{\tilde{q}_2}) \Big]  \non\\
&&  +  \szt{q} \; \frac{\mg}{m_q} \; \Big[ B_0(m_q^2,\mg,m_{\tilde{q}_1})
           -B_0(m_q^2,\mg,m_{\tilde{q}_2}) \Big]
\eeq
and 
\beq
 \delta \tht{q} =    \frac{1}{m_{\tilde{q}_1}^2-m_{\tilde{q}_2}^2}
  \bigg[ 4 \mg\, m_q \czt{q}\; B_0(m_{\tilde{q}_1}^2,m_q,\mg) 
       + \czt{q}\, \szt{q}\; 
         \Big( A_0(m_{\tilde{q}_2})-A_0(m_{\tilde{q}_1}) \Big) \bigg]
\eeq
where again $i'=i-3$ in eq.~(\ref{sqm_ct}) and the functions $B_0'$ and $B_1$ are
defined as 
\beq
B_0'( q^2, m_1, m_2) &=& \frac{\partial }{\partial q^2} \, B_0 (q^2, m_1, m_2)
\non \\
B_1( q^2, m_1, m_2) &=& \frac{1}{2q^2}\bigg[A(m_1)-A(m_2) +
      (m_2^2-m_1^2-q^2)B_0 (q^2, m_1, m_2)\bigg]
\eeq

The sum of all counterterm corrections is then given by
\beq
\Delta^{\rm CT}_{H^\pm  ij} & = &  \frac{1}{2} G_{\Phi ij} 
   \Big(\delta Z_{\tilde{t}_i} + \delta Z_{\tilde{b}_j} \Big)
    + \frac{\partial \,G_{\Phi ij}}{\partial\, m_t}\, \delta m_t
    + \frac{\partial \,G_{\Phi ij}}{\partial\, m_b}\, \delta m_b \non\\
&&  + \frac{\partial \,G_{\Phi ij}}{\partial\, A_t}\, \delta A_t
    + \frac{\partial \,G_{\Phi ij}}{\partial\, A_b}\, \delta A_b\;
    + \frac{\partial \,G_{\Phi ij}}{\partial\, \tht{t}}\, \delta \tht{t}
    + \frac{\partial \,G_{\Phi ij}}{\partial\, \tht{b}}\, \delta \tht{b}
\eeq
for the charged Higgs boson, whereas for neutral Higgs decays it reads
\beq
  \Delta^{\rm CT}_{\Phi ij} & = &  \frac{1}{2} G_{\Phi ij} 
   \Big(\delta Z_{\tilde{q}_i} + \delta Z_{\tilde{q}_j} \Big)
    + \frac{\partial \,G_{\Phi ij}}{\partial\, m_q}\, \delta m_q
    + \frac{\partial \,G_{\Phi ij}}{\partial\, A_q}\, \delta A_q
    + \frac{\partial \,G_{\Phi ij}}{\partial\, \tht{q}}\, \delta \tht{q}
\eeq

Note that only the wave--function counterterm involves infrared divergences 
and the gluon mass $\lambda$ can be set to zero everywhere except in the last 
term of eq.~(\ref{d_Zsq}). \s

In Ref.~\cite{R8}, the renormalization of the quark and squark masses
has been performed in the $\overline{\rm DR}$ scheme, i.e. by subtracting
the poles and the related constants in the dimensional reduction scheme; 
in addition there was no explicit counterterm for the  trilinear squark
coupling $A_q$. The renormalization scheme adopted here is more convenient,
since the conditions we used for the masses allow for a straightforward
physical interpretation. 

\subsubsection*{3.3 Real corrections}

Finally, the real corrections with an additional gluon in the final state
need also to be included. In agreement with Ref.~\cite{R8}, we obtain 
\beq
\Delta^{R}_{\Phi ij} = \frac{8 M_\Phi^2 G_{\Phi ij} }{\lambda^{1/2}(M_\Phi^2,
m_{\tilde{q}_i}^2, m_{\tilde{q}'_j}^2)}
\Big[ (M_{\Phi}^2-m_{\tilde{q}_i}^2-m_{\tilde{q}'_j}^2) I_{12}
                 - m_{\tilde{q}_i}^2 I_{11}
                 - m_{\tilde{q}'_j}^2 I_{22} - I_1 - I_2\Big]
\eeq
where the phase space integrals $I_{ij}$ and $I_{i,j}$ are defined as
\beq
I_{ij}& = &\frac{1}{\pi^2}\int \frac{d^3k_{g}}{2E_{g}} \frac{d^3k_{i}}{2E_{i}}
\frac{d^3k_{j}}{2E_{j}} \frac{ \delta^4 (k_\Phi-k_i-k_j -k_g)}{ (2 k_{g}.k_i)  
(2 k_{g} . k_j)} \nonumber\\
I_{i,j}& =&\frac{1}{\pi^2}\int \frac{d^3k_{g}}{2E_{g}} \frac{d^3k_{i}}{2E_{i}}
\frac{d^3k_{j}}{2E_{j}} \frac{\delta^4(k_\Phi-k_i-k_j-k_g)}{ (2 k_{g} . k_{i,j}) }
\eeq
with $k_g, k_{i,j}$  the four-momenta of the gluon and the squarks
$\tilde{q}_{i,j}$. In terms of Dilogarithms, this gives 
\beq
\Delta^R_{\rm \Phi ij} & = & \frac{2 G_{\Phi ij}}{ \lambda_{ij}^{1/2} } \Big\{
(M_{\Phi}^2-m_{{\tilde q}_i}^2- m_{{\tilde q'}_j}^2)  
\Big[ -2 \log \left( \frac{\lambda M_{\Phi}m_{{\tilde q}_i} m_{{\tilde q'}_j} 
} {\lambda_{ij} } \right) \log\beta_0 +  2 \log^2\beta_0  \non \\
&& -\log^2 \beta_1 - \log^2\beta_2 +     2 {\rm Li_{2}} (1-\beta_0^2)-{\rm 
Li_{2}} (1-\beta_1^2)- {\rm Li_{2}} (1-\beta_2^2) \Big] \non \\ 
&& + 2 \lambda_{ij}^{1/2} \log \left( \frac{\lambda M_{\Phi}
m_{{\tilde q}_i} m_{{\tilde q'}_j }}  {\lambda_{ij}} \right) + 
4 \lambda_{ij}^{1/2}  + ( 2 M_{\Phi}^2 +   m_{{\tilde q}_i}^2 + 
m_{{\tilde q'}_j}^2 ) \log\beta_0 \non \\ 
&& +( M_{\Phi}^2 + 2 m_{{\tilde q'}_j}^2 ) \log\beta_2  
+ (M_{\Phi}^2 + 2 m_{{\tilde q}_i}^2 ) \log\beta_1 \big\}
\eeq
with $\lambda_{ij} \equiv \lambda(M_\Phi^2, m_{{\tilde q}_i}^2 ,
m_{{\tilde q'}_j}^2)$ and 
\begin{eqnarray}
\beta_0 = \frac{M_{\Phi}^2- m_{{\tilde q}_i}^2 - m_{{\tilde q'}_j}^2 
+\lambda_{ij}^{1/2} }{2 m_{{\tilde q}_i} m_{{\tilde q'}_j} } \hspace*{4cm}  
\non \\ 
\beta_{1} = \frac{M_{\Phi}^2- m_{{\tilde q}_i}^2 + m_{{\tilde q'}_j}^2 
-\lambda_{ij}^{1/2} }{2 M_{\Phi}  m_{{\tilde q'}_j} } \ \ , 
 \ \ {\rm  and } \ \ \beta_{2} = \frac{M_{\Phi}^2+ m_{{\tilde q}_i}^2 - 
m_{{\tilde q'}_j}^2 -\lambda_{ij}^{1/2} }{2 M_{\Phi}  m_{{\tilde q}_i} }
\end{eqnarray}

\subsection*{4. Numerical Results} 

In the numerical analysis of the QCD corrections we have chosen $m_t=175$ GeV
and $m_b=5$ GeV for the top and bottom masses, $s_W^2=0.23$ for the electroweak
mixing angle, and $\alpha_s = 0.12$ for the strong coupling constant. In 
Figs.~1--4, $m_{\tilde{t}_1}$ always denotes the mass of the lightest scalar 
top quark. The Higgs masses are fixed by $M_A=600$ GeV, and for $\tan\beta$ 
we choose the value $\tan\beta=1.6$. We will always assume that the 
left-- and right--handed scalar masses are equal, $m_{\tilde{t}_L}=
m_{\tilde{t}_R}=m_{\tilde{b}_L}=m_{\tilde{b}_R}$. \s

The dominating decay modes of the neutral Higgs bosons are the decays
$\Phi \ra \tilde{t}_i \tilde{t}_j$ into scalar top quarks. In order to
visualize the effect of the QCD corrections, Figs.~2 and 3 display the
decay widths of the heavy CP--even Higgs boson $H$ and the pseudoscalar Higgs
boson $A$ 
into top squark pairs in Born approximation (solid lines) and including the 
QCD corrections for two values of the gluino mass: $m_{\tilde{g}}=200$ GeV 
(dashed lines) and $m_{\tilde{g}}=1$ TeV (dotted lines). \s

Fig.~2a reflects 
the situation for unmixed top squarks, where the choice $A_t = -\mu\,\ctg
\beta$ yields a diagonal stop mass matrix eq.~(\ref{sqmass_matrix}). Fig.~2b 
refers to maximal mixing $\theta_t \simeq -\pi/4$ in the scalar top sector, 
with $A_t=250$ GeV. $\mu=-300$ GeV is a common input for Figs.~2a,b. The 
decay widths are significantly larger for the case of mixing, being further
increased by large QCD corrections up to nearly 50\%, whereas in the unmixed
case the QCD corrections decrease the Born width significantly for the major
part of the $\tilde{t}_1$ mass range. Only close to the phase space boundary,
the higher order contribution is positive. \s

The non--diagonal decay mode $H \ra \tilde{t}_1 \tilde{t}_2$, shown in 
Fig.~2c for $\mu=100$ GeV and $A_t=150$ GeV, has a similar signature as 
the diagonal one in Fig.~2b, but it is suppressed by several orders of 
magnitude and is thus of less interest. \s

For the CP--odd $A$ boson, the non--diagonal decay is the only allowed decay
mode. The width is comparable in size to that of the diagonal $H$ boson decay.
We discuss the impact of QCD corrections in terms of two examples:
Fig.~3a corresponds to the parameters $\mu=-300$ GeV, $A_t=300$ GeV.
Here the corrections show a similar pattern as for the diagonal decay mode
of the $H$ with maximal mixing, Fig. 2b, having a milder dependence on the
gluino mass. For other values of the $\mu$, $A_t$ parameters where the lowest
order $A$ width is smaller, the effect of the QCD corrections can be much
more dramatic, as shown in Fig.~3b for $\mu=100$ GeV, $A_t=250$ GeV. Depending
crucially on the gluino mass, the QCD loop contribution can be either positive
($m_{\tilde{g}}=200$ GeV) or negative ($m_{\tilde{g}}=1$ TeV), in both cases
up to the order of 50\%. The kink in Fig.~3b corresponds to the threshold
$m_{\tilde{t}_2}=m_{\tilde{g}}+m_t$ in the $\tilde{t}_2$ wave function
renormalization. \s

For completeness, we also present numerical results for the decay of the charged
Higgs boson $H^+ \ra \tilde{t}_1 \tilde{b}_1$ in order to demonstrate the impact
of the QCD contributions. This decay was already studied earlier in the
literature \cite{R8} using a $\overline{\rm DR}$ renormalization scheme. As
for the decay of the heavy CP--even Higgs boson $H$, we consider the two 
situations of unmixed top squarks (Fig.~4a) obtained by $A_t=-\mu \,\ctg
\beta$, $A_b=-\mu\;\tan\beta$, and for the case of mixing with $A_t=A_b=-100$ 
GeV (Fig.~4b). Both  Figs.~4a,b are for $\mu=200$ GeV. The QCD corrections are
large and negative, independent of top squark mixing. They can decrease the partial
decay width by almost an order of magnitude. \s

As seen previously the magnitude of the QCD correction strongly depends on
the gluino mass. In Fig.~5a we show the dependence of the $H$ boson 
decay width on $m_{\tilde{g}}$, which is treated here as an independent
parameter. The two curves correspond to the case of unmixed top squarks
(dotted) and stop mixing (dashed--dotted) with $A_t=250$ GeV. The other
parameters are fixed according to $\mu=-300$ GeV, $m_{\tilde{t}_L}=150$ GeV
and the solid and dashed lines indicate the respective widths in the born
approximation. Whereas for the unmixed case the QCD contributions
decrease the width for large $m_{\tilde{g}}$, the decay width increases with
$m_{\tilde{g}}$ for the mixing case. The asymptotic dependence on
$m_{\tilde{g}}$ is logarithmic; all terms linear in $m_{\tilde{g}}$ cancel
in the final result. Such a logarithmic behavior has been first observed 
and discussed in the case of the decay of squarks into quarks and photinos
\cite{R17}. \s

The asymptotic dependence on $m_{\tilde{g}}$ can be cast into a simple form.
In the notation and normalization of eqs.~(7--10), the decay amplitude
$G_{A\tilde{t}_1\tilde{t}_2}$ for the process $A \ra \tilde{t}_1 
\overline{\tilde{t}}_2$ gets an additive gluino mass dependent
contribution of the type
\beq
\Delta G_{A \tilde{t}_1 \tilde{t}_2} &\simeq& \frac{\,m_t}{\sqrt{2}}
\left[(A_t-M^t_{LR})\,\ctg\beta-\mu\right]\,\frac{4\,\alpha_s}{3\,\pi}
\log(m_{\tilde{g}}) \non\\
&=& -\frac{m_t\,\mu}{\sqrt{2}\sin^2\beta}\,\frac{4\,\alpha_s}{3\,\pi}
\log(m_{\tilde{g}})\;.
\eeq
For the $H$ boson decays, the structure is slightly more complicated, but 
still very compact: The coupling matrix $G_{H \tilde{t} \tilde{t}}$ in eq.
(\ref{couplingmatrix}) gets the following additive contribution:
\beq
\Delta G_{H \tilde{t} \tilde{t}} & = & \frac{4\,\alpha_s}{3\,\pi}
\log(m_{\tilde{g}})\,\Delta C_{H \tilde{t} \tilde{t}}
\eeq
with the matrix
\beq
\Delta C_{H \tilde{t} \tilde{t}} & = & 2\,(R^{\tilde{t}})^T
   C_{H \tilde{t} \tilde{t}} R^{\tilde{t}} \non \\
& &\hspace{-0.5cm}
-\frac{1}{\sqrt{2}}\,(R^{\tilde{t}})^T
\left(
\begin{array}{cc}
8\,r_1^t\,m_t^2                    & m_t [(A_t+M_{LR}^t)r_1^t+\mu\,r_2^t] \\ 
m_t [(A_t+M_{LR}^t)r_1^t+\mu\,r_2^t]  & 8\,r_1^t\,m_t^2
\end{array}
\right)
                      R^{\tilde{t}}
\eeq
where $C_{H \tilde{t} \tilde{t}}$ is the matrix in
eq.~(\ref{h0couplings}). For the special case $H \ra \tilde{t}_1
\overline{\tilde{t}}_1$ we obtain:
\beq
\Delta C_{H \tilde{t}_1 \tilde{t}_1} & = & 2\, C_{H \tilde{t}_1 \tilde{t}_1}
-\frac{1}{\sqrt{2}}\bigg\{8\,m_t^2 r_1^t + 2\,\st{t}\ct{t}\,
m_t [(A_t+M_{LR}^t)\,r_1^t+\mu\,r_2^t] \bigg\} \non \\
& \simeq & 2\,C_{H \tilde{t}_1 \tilde{t}_1}
+\frac{1}{\sqrt{2}}\bigg\{8\,m_t^2 \ctg\beta + 2\,\st{t}\ct{t}\,
m_t [(A_t+M_{LR}^t)\ctg\beta-\mu] \bigg\}
\eeq
From this expression, the different behaviour of the $m_{\tilde{g}}$--dependence
for the mixed and unmixed case in Fig.~5 can be qualitatively understood.
It should be noted that a strong $m_{\tilde{g}}$--dependence for intermediate
values of $m_{\tilde{g}}$ is introduced via the finite part of the 
vertex diagram with $q$ and $\tilde{g}$ in the internal lines (Fig.~1b). 
It decouples for large $m_{\tilde{g}}$, but the decoupling is very slow. \s

This gluino vertex diagram is also responsible for a finite pseudoscalar 
decay width in case that the tree level coupling of $A$ to $\tilde{q}_1 
\tilde{q}_2$ is zero, i.e. for correlating $\mu$ and $A_q$ such that the 
entries in eq.~(\ref{a0couplings}) vanish. Together with the counterterm 
from the renormalization of the trilinear coupling $A_q$, eq.~(\ref{AqCT}), 
the 1--loop amplitude is ultraviolet and infrared finite and gives rise to 
a loop--induced finite decay width. This situation is shown in Fig.~5b for
$A^0 \ra \tilde{t}_1 \overline{\tilde{t}}_2$ for the parameters $A_t=\mu\,\tan\beta$,
$\mu=\pm 100$ GeV and $m_{\tilde{t}_L}=200$ GeV as a function of the gluino 
mass. The strong dependence on $m_{\tilde{g}}$ is to a large extent due to 
the finite part of the gluino vertex diagram, eq. (\ref{gluinovertex}).

\subsection*{5. Squark decays into Higgs Bosons}

For completeness, let us also discuss the decays of scalar quarks into 
lighter squarks and Higgs bosons. In practice this situation can occur
when there is a large mass splitting between the heaviest and the lightest
top squarks where the decay
\beq 
\tilde{t}_2 \ra \tilde{t}_1 \ + \ h/A/H 
\eeq
can occur, and between the stop and sbottom squarks where the decays
\beq
\tilde{t}_2 \ra H^+ \, + \, \tilde{b}_1 \ \ {\rm or} \ \ \tilde{b}_1 
\ra H^- \, + \, \tilde{t}_1 
\eeq
can be kinematically allowed. In the previous equation, $\tilde{b}_1$ is 
the lightest bottom squark which is identical to the left--handed 
$\tilde{b}$ squark in case of no mixing in the sbottom sector; decays 
of the heaviest bottom squark into neutral Higgs bosons can also be 
possible for large $\tan \beta$ values where the $\tilde{b}_1$--$\tilde{b}_2$ 
mass splitting and the sbottom mixing angle can be large enough. These 
decays have been discussed at tree--level in Ref.~\cite{R10a} for instance. \s

The partial decay width of a squark $\tilde{q}_i$ to a Higgs boson $\Phi$ and 
a lighter squark $\tilde{q}_j'$ is given by a relation similar to eq.~(\ref{phiborn})
but without the color factor
\beq
\Gamma^0 (\tilde{q}_i \ra \Phi\, \overline{\tilde{q}}'_j) = \frac{G_F }{4
\sqrt{2} \pi \, m_{\tilde{q}_i}^3 } \, \lambda^{1/2} (m_{\tilde{q}_i}^2, 
M_{\Phi}^2, m_{\tilde{q}'_j}^2) \; \left( \; G_{\Phi i j} \; \right)^2
\eeq
with the coupling $ G_{\Phi i j}$, with $\Phi=H,A$ and $H^\pm$  given as in 
section 2. The couplings of the lightest CP--even Higgs boson $h$ to 
squark pairs  can be obtained from those of the $H$ boson by simply replacing 
$\alpha$ by $\alpha-\pi/2$. \s

The QCD corrections to the decay width are given by the Feynman diagrams shown 
in Fig.~1 with now $\tilde{q}_i$ and $\Phi$ being in the initial and final 
state respectively. The QCD corrected decay width can be written as
 \beq
\Gamma^1 (\tilde{q}_i \ra \Phi \, \overline{\tilde{q}}'_j) = \frac{G_F}
{6 \sqrt{2} \pi \, m_{\tilde{q}_i}^3} \, \frac{\alpha_s}{\pi} \,
\lambda^{1/2} (m_{\tilde{q}_i}^2, 
M_{\Phi}^2, m_{\tilde{q}'_j}^2)  \; G_{\Phi i j} \;  \Delta_{\Phi ij} 
\eeq
where, similarly to eq.~(\ref{virtcorr}), the correction factor $\Delta_{\Phi ij}$ is 
given by
\beq
\Delta_{\Phi ij}=\Delta^V_{\Phi ij}+\Delta^{\rm CT}_{\Phi ij}
+\Delta^R_{\Phi ij}\;.
\eeq
The virtual corrections $\Delta^V_{\Phi ij}$ and the corresponding 
counterterm $\Delta^{\rm CT}_{\Phi ij}$ are given by the same expressions 
as in section 3. In the real corrections, however, the role of $\tilde{q}_i$
and the Higgs boson $\Phi$ have to be interchanged:
\beq
\Delta^R_{\Phi ij} = \frac{8 m_{\tilde{q}_i}^2 G_0^{\Phi ij} } 
{ \lambda^{1/2} ( m_{\tilde{q}_i}^2, M_{\Phi}^2, m_{\tilde{q}'_j}^2) }     \,
              \Big[ (M_{\Phi}^2-m_{\tilde{q}_i}^2-m_{\tilde{q}'_j}^2) I_{02}
                 - m_{\tilde{q}_i}^2 I_{00}
                 - m_{\tilde{q}'_j}^2 I_{22} - I_0 - I_2 \Big]
\eeq
where the functions $I_{lk}$ \cite{den} have arguments  
\beq
I_{lk} \equiv I_{lk}(m_{\tilde{q}_i}, m_{\tilde{q}'_j}, M_{\Phi},\lambda)  
\eeq

Fig.~6 shows the partial decay widths of the heaviest top squark $\tilde{t}_2$
into its lighter partner $\tilde{t}_1$ and the lightest neutral CP--even Higgs 
boson $h$ (6a) and the pseudoscalar Higgs boson $A$ (6b) as a function of the 
lightest top squark mass. Again, the solid lines show the tree--level decay 
widths, while the dashed and dotted lines show the QCD corrected widths for 
$m_{\tilde{g}} =200$ GeV and $m_{\tilde{g}} =1$ TeV, respectively. To have the 
$\tilde{t}_2$--$\tilde{t}_1$ mass splitting large enough while leaving the 
decay rates sizeable, 
we relax the condition $m_{\tilde{t}_L} = m_{\tilde{t}_R}$ for the stop mass 
parameters. We chose $\mu=-300$ GeV, $\tb=1.6$, $A_t =300$ GeV and 
$m_{\tilde{t}_R}=500$ GeV; $m_{\tilde{t}_L}$ is then varied from 200 to
430 GeV. \s

In Fig.~6a, the pseudoscalar Higgs boson mass is taken to be $M_A=400$ GeV, 
leading
to a value $M_h \simeq 75$ GeV. This value slightly varies with 
$m_{\tilde{t}_L}$, since we include the large radiative corrections 
in the Higgs sector which increase with $m_t^4$ and with the logarithm of
$m_{\tilde{t}_L}m_{\tilde{t}_R}$. The $\tilde{t}_2 \ra h \tilde{t}_1$
decay width is about 0.1 GeV for small $\tilde{t}_1$ masses and
the QCD corrections increase the decay width by an amount of ${\cal 
O}(10\%)$. \s 

In Fig.~6b, the pseudoscalar mass is chosen to be $M_A=100$ GeV. The
$\tilde{t}_2 \ra \tilde{t}_1 A$ decay width is an order of magnitude 
larger than in the previous case, but the QCD corrections are now 
negative, decreasing the partial widths by 20\%. Again, the kinks 
in Figs.~6 are due to the opening of the $\tilde{t}_2 \ra t+\tilde{g}$
threshold. Above this value, this decay becomes by far the dominant one
compared to the decays into Higgs bosons. 

\subsection*{6. Conclusions}

We have calculated the SUSY QCD corrections to the decays of the MSSM 
heavy neutral and charged Higgs bosons into scalar quark pairs, including 
the mixing in the squark sector. A special attention has been paid to the case
of stop and sbottom decays which can be the dominant decay channels in a large 
area of the MSSM parameter space. The QCD corrections turn out to be quite 
substantial, enhancing or suppressing the decay widths in Born approximation
by amounts up to 50\% and in some cases more. The QCD corrections depend 
strongly on the
gluino mass; however, for large gluino masses, the QCD correction is
only logarithmically dependent on $m_{\tilde{g}}$. Contrary to the case
of Higgs decays into light quarks, these QCD corrections cannot be absorbed 
into running squark masses since the latter are expected to be of the same 
order of magnitude as the Higgs boson masses. \s

We have also calculated the QCD corrections to the decays of heavy squarks 
into lighter squarks and Higgs bosons. Cases of interest are for instance the
channels $\tilde{t}_2 \ra \tilde{t}_1+h/A$. The QCD corrections are at the
level of a few ten percent and can enhance or suppress the tree--level
decay widths. In both cases, Higgs decays into scalar quarks and squark 
decays into Higgs bosons, the QCD corrections are therefore important 
and should be included when discussing these decays. 

\bigskip

After completion of this work, another paper appeared \cite{R18} which deals 
with the decays of the MSSM Higgs bosons into squark pairs. 
It is an extension of the earlier work \cite{R8} for the $H^+$ decay, but the 
discussion is performed also in the on--shell scheme. The difference to 
our renormalization scheme is only a slightly different treatment of the 
squark mixing angle renormalization, which results in a minor finite shift 
in the mixing angle counterterm. We have checked that all analytical 
expressions agree with ours. Numerical comparisons also show good agreement;
in our case we show that the correction can strongly decrease 
the decay widths and determine the origin of the large corrections. The
decays of heavy squarks to lighter squarks and Higgs bosons has not been 
discussed in Ref.~\cite{R18}. 

\bigskip

{\bf Acknowledgments}: We thank Roland H\"opker for his collaboration
in the early stage of this work.

\newpage

\nn \subsection*{Figure Captions}
\begin{itemize}
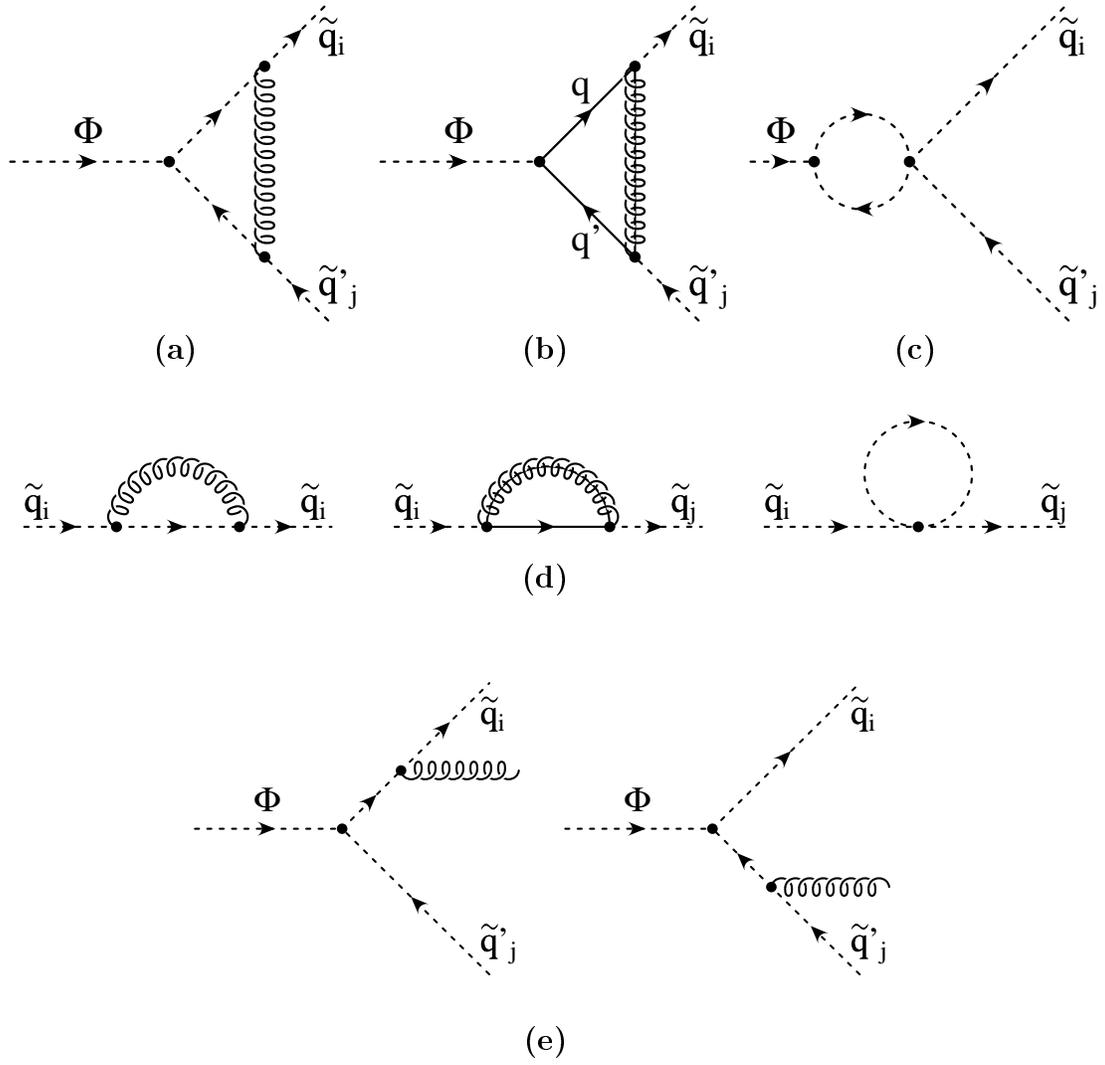
\item[{\bf Fig.~1:~}]
Feynman diagrams relevant for the ${\cal O}(\alpha_s)$ QCD corrections
to the decay of a Higgs boson $\Phi$ into scalar quark pairs. 
(a): gluon exchange, (b) gluino exchange, (c) mixing diagrams,
(d) quartic squark interaction, (e) self-energy and vertex 
counterterms, (f) real corrections. 
\vspace{-1mm}
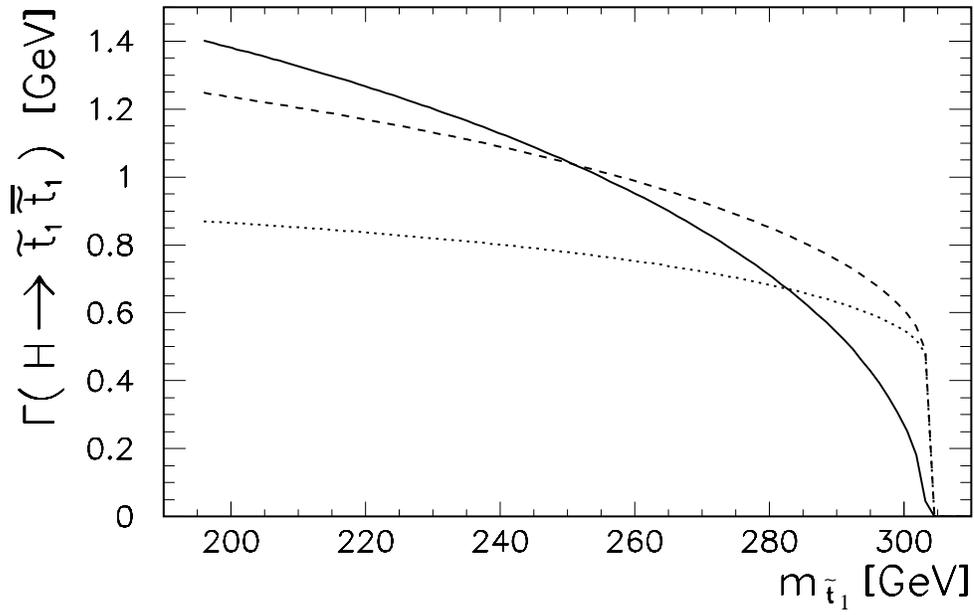
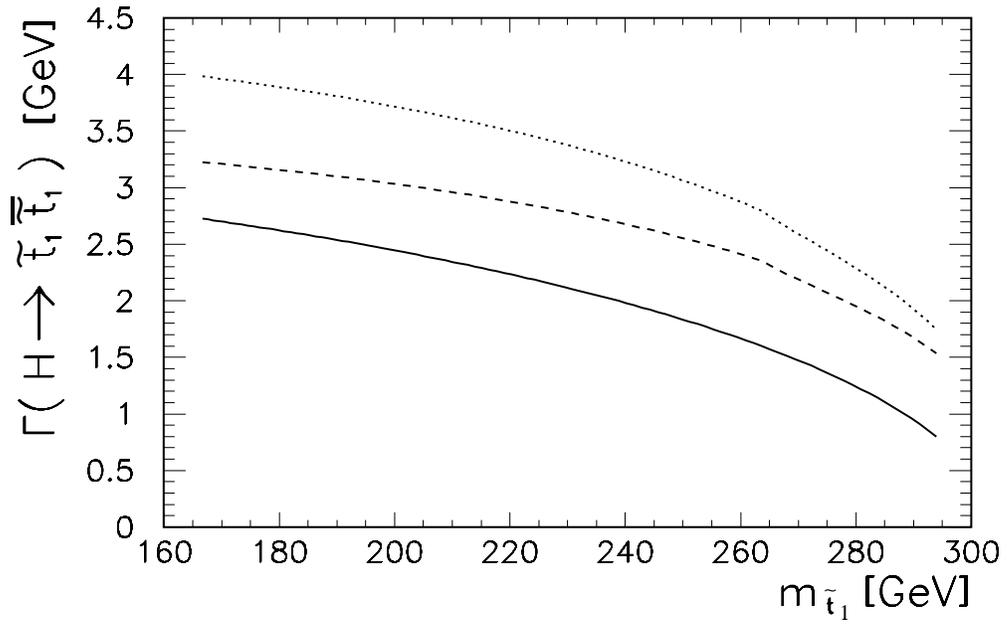
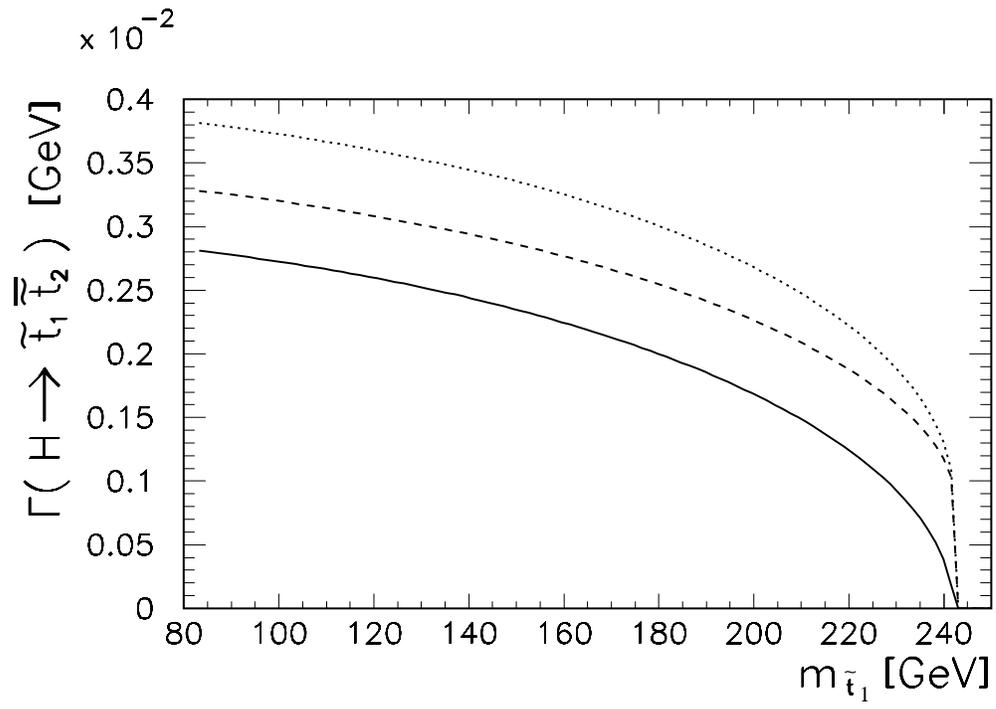
\item[{\bf Fig.~2:~}]
Partial widths [in GeV] for the decays of the neutral CP--even Higgs boson
into top squark pairs, $H \ra \tilde{t}_i \tilde{t}_j$, as a function of 
of $m_{\tilde{t}_1}$. The Higgs mass is fixed by $M_A=600$ GeV and $\tb=1.6$;
$\mu = -300$ GeV, $A_t = -\mu \,\ctg\beta$ (a); $\mu = -300$ GeV and
$A_t = 250$ GeV (b); $\mu = 100$ GeV and  $A_t = 150$ GeV (c). 
The solid lines are for the partial widths in the
Born approximation, while  the dashed and dotted lines are for the partial
widths including QCD corrections for $\mg=200$ GeV and 1 TeV respectively. 
\vspace{-1mm}
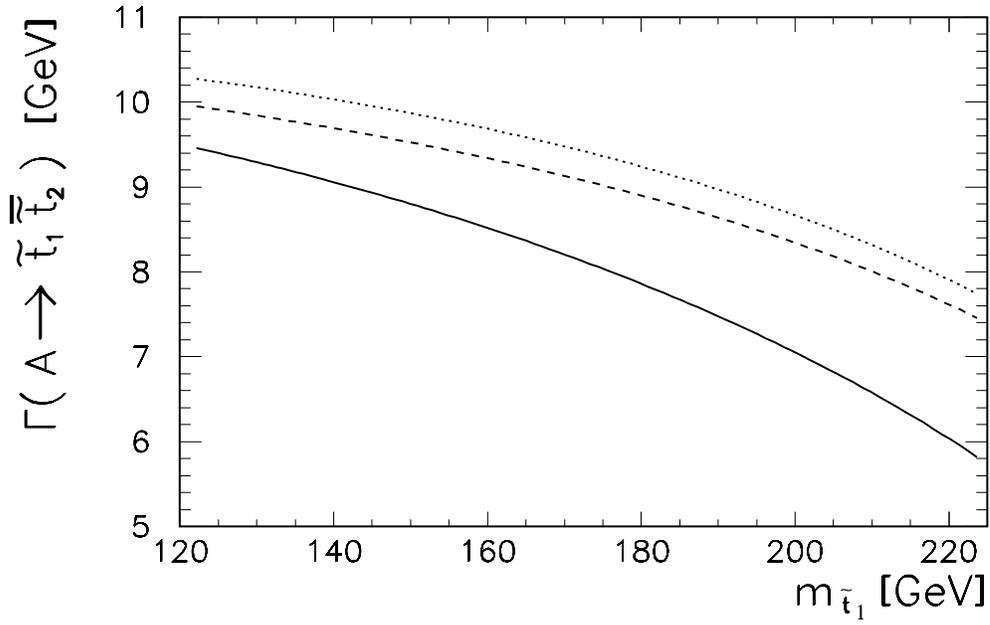
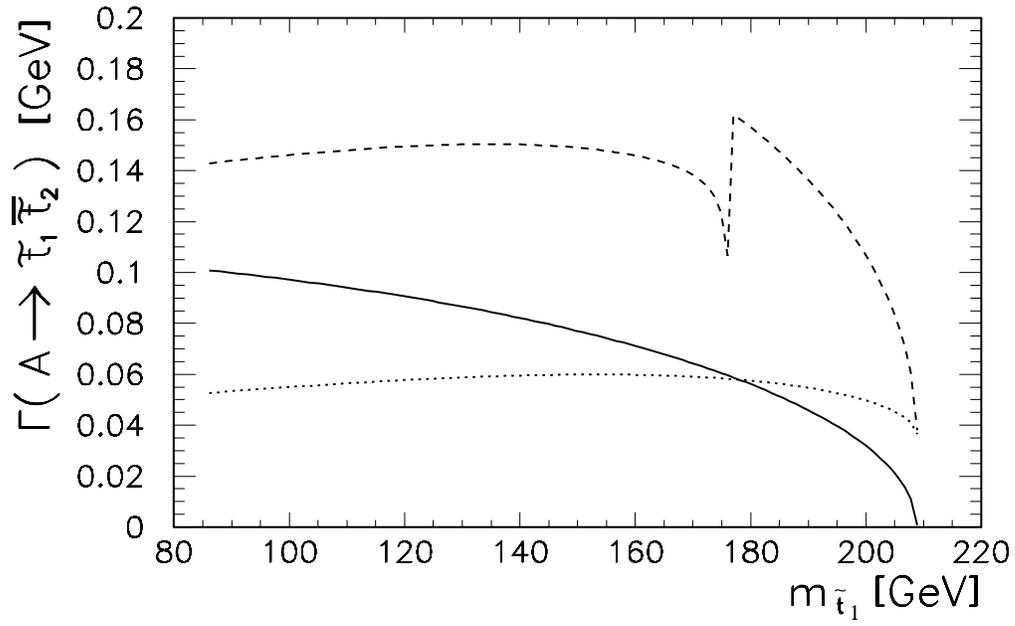
\item[{\bf Fig.~3:~}]
Partial widths [in GeV] for the decays of the neutral CP--odd Higgs boson
into top squarks, $A \ra \tilde{t}_1 \tilde{t}_2$, as a function of
$m_{\tilde{t}_1}$. The Higgs masses are fixed by $M_A = 600$ GeV, $\tb=1.6$;
$\mu = -300$ GeV and $A_t = 300$ GeV (a); $\mu = 100$ GeV and $A_t = 250$ GeV (b). The solid lines are for the partial widths in the Born 
approximation, while the dashed and dotted lines are for the partial widths 
including QCD corrections for $\mg=200$ GeV and 1 TeV respectively. 
\vspace{-1mm}
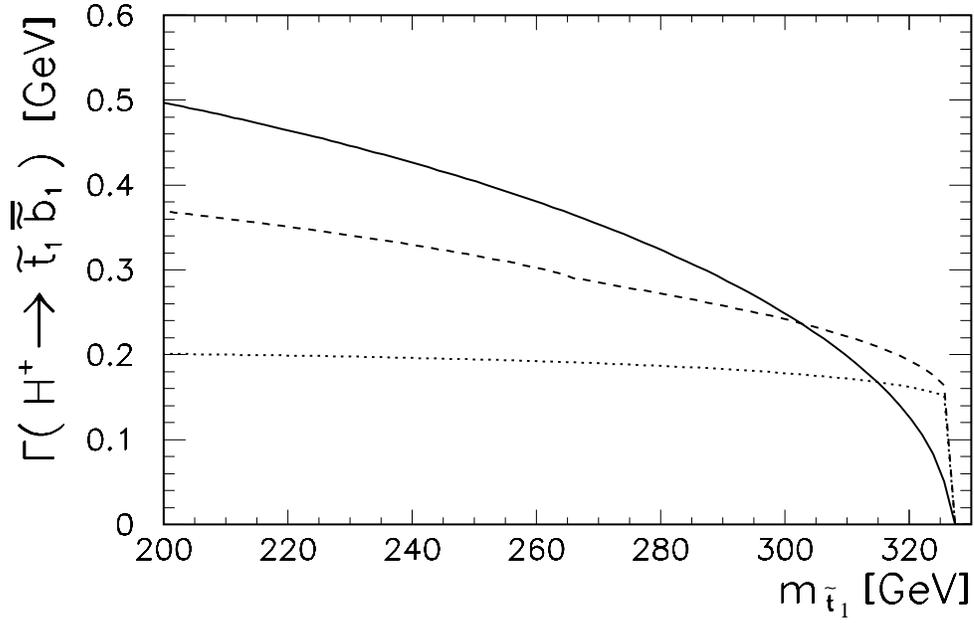
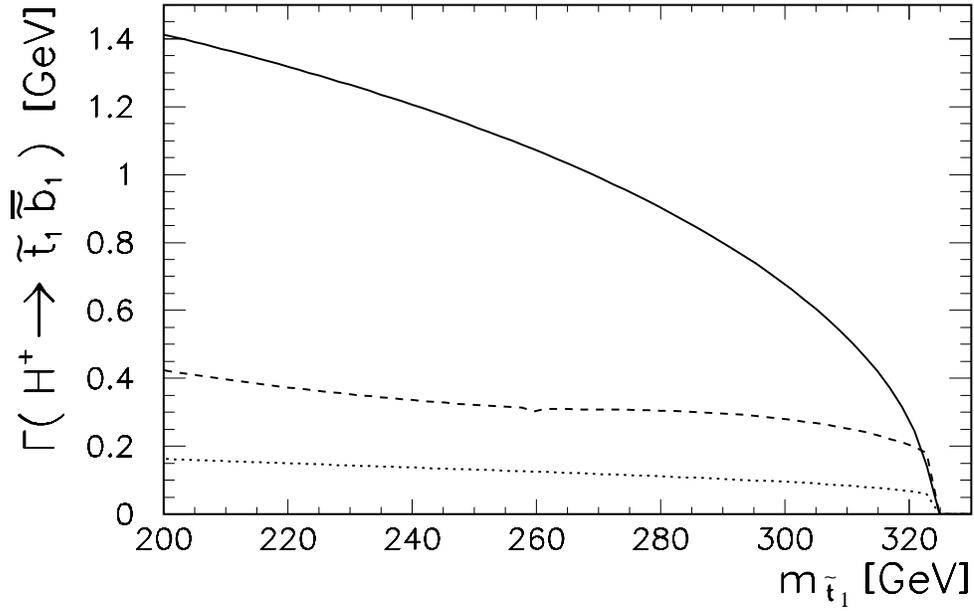
\item[{\bf Fig.~4:~}]
Partial widths [in GeV] for the decays of the charged Higgs boson into
the lightest top and bottom squarks $H^\pm \ra \tilde{t}_1 \tilde{b}_1$ 
as a function of $m_{\tilde{t}_1}$. The Higgs mass is fixed by $M_A =
600$ GeV and $\tb=1.6$; $\mu = 200$ GeV, $A_t = -\mu\,\ctg\beta$ and
$A_b = -\mu\,\tan\beta$ (a) and $\mu = 200$ GeV and $A_t = A_b = -100$ GeV (b).
The solid lines are for the partial widths in the 
Born approximation, while the dashed and dotted lines are for the partial 
widths including QCD corrections for $\mg=200$ GeV and 1 TeV respectively. 
\vspace{-1mm}
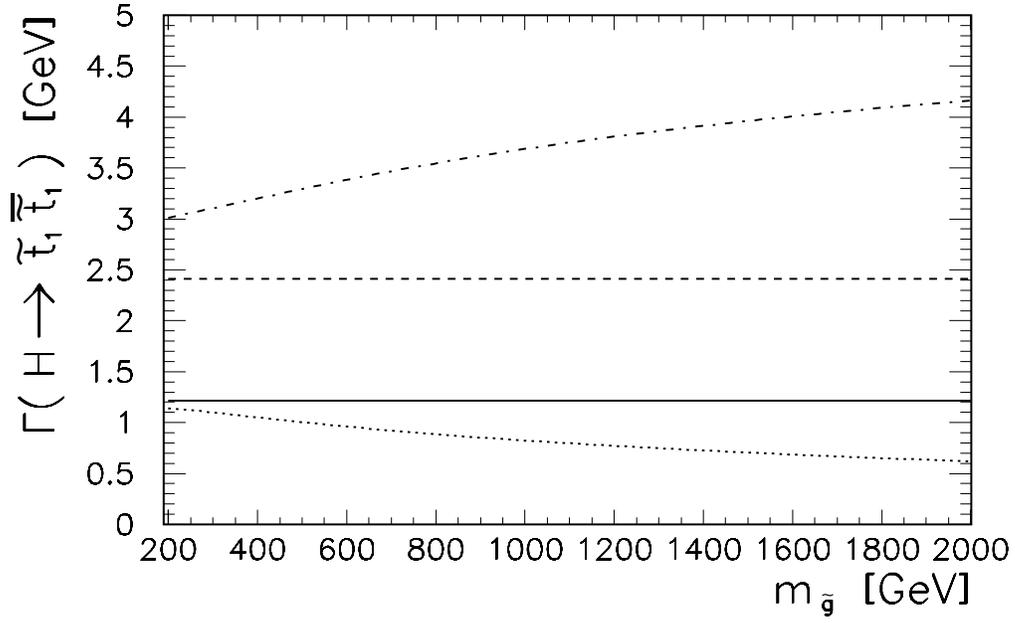
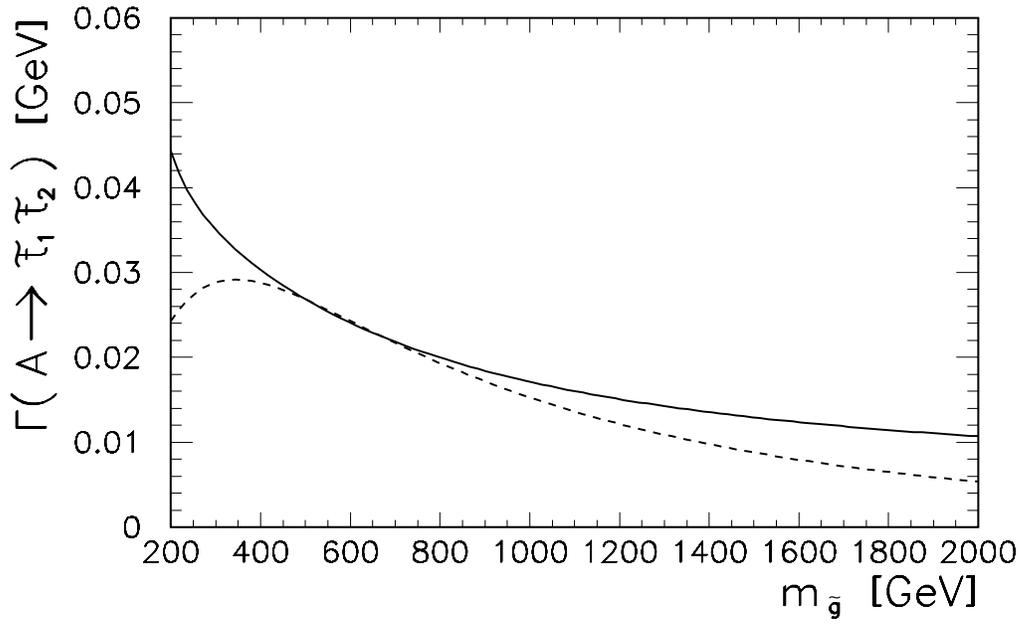
\item[{\bf Fig.~5:~}]
Partial width [in GeV] for the decay of the neutral CP--even Higgs boson
$H$ into the lightest top squarks (a) and of the neutral CP--odd Higgs boson
$A$ into top squarks, as a function of the gluino mass. The Higgs boson masses 
are fixed by $M_A = 600$ GeV and $\tb=1.6$. In (a) we take $\mu = -300$ GeV, 
$m_{\tilde{t}_L} = 150$ GeV and $A_t = -\mu\,\ctg\beta$ (dotted) and $\mu = 
-300$ GeV, $m_{\tilde{t}_L} = 150$ GeV and $A_t = 250$ GeV (dashed--dotted).
The solid and dashed lines corrrespond to the respective born widths.
In (b): $A_t$ is fixed to make the lowest order width vanish,
$A_t = \mu \, \tan\beta$ and the stop masses are fixed by $m_{\tilde{t}_L} 
= 200$ GeV; $\mu = 100$ GeV (solid) and $\mu = -100$ GeV (dashed).
\vspace{-1mm}
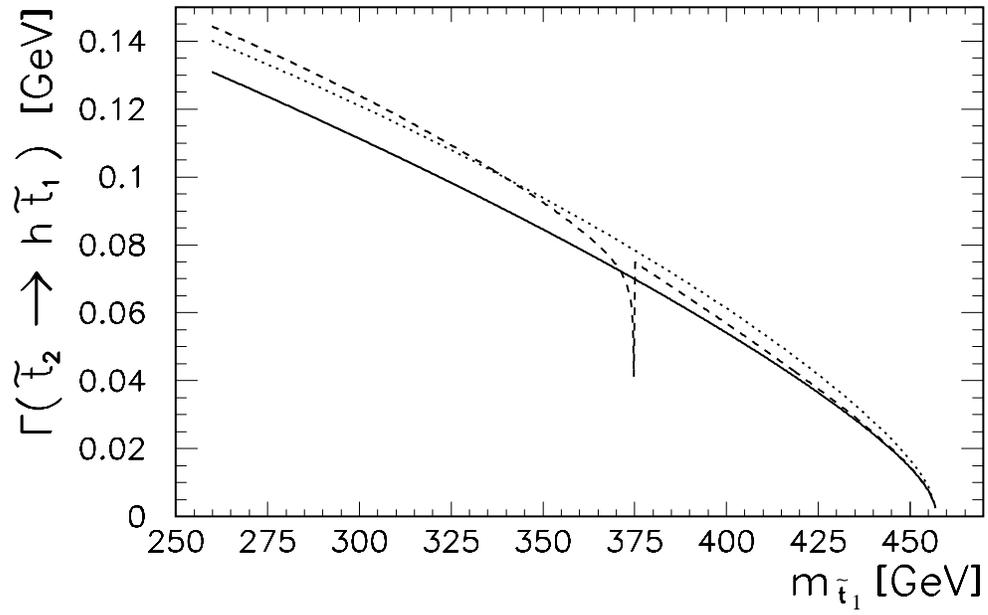
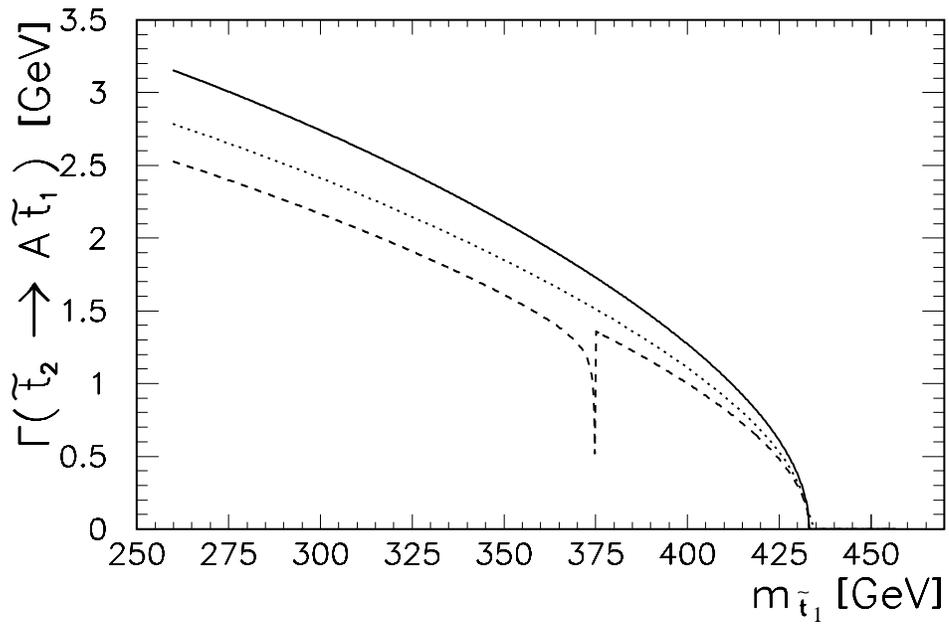
\item[{\bf Fig.~6:~}]
Partial decay widths of the heaviest top squark $\tilde{t}_2$ into 
$\tilde{t}_1$ and the lightest neutral CP--even Higgs boson $h$
(a) and the pseudoscalar Higgs boson $A$ (b) as a function of $m_{\tilde{t}_1}$.
The solid lines show the tree--level decay widths, while 
the dashed and dotted lines show the QCD corrected widths for $m_{\tilde{g}} 
=200$ GeV and 1 TeV, respectively. $\mu=-300$ GeV, $\tb=1.6$, $A_t =300$ GeV 
and $m_{\tilde{t}_R}=500$ GeV while $m_{\tilde{t}_L}$ is varied from
200 to 430 GeV. For the Higgs masses we take $M_A=400$ GeV (a) and
$M_A=100$ GeV (b). 
\end{itemize}


\newpage 

\begin{thebibliography}{99}

\bibitem{R1} For reviews on Supersymmetry see: H. P. Nilles, Phys. Rep. 
117 (1985) 1; \\
P. Nath, R. Arnowitt and A. Chamseddine, {\it Applied N=1
Supergravity}, ICTP series in Theoretical Physics, World Scientific,
Singapore, 1984; \\ Haber and G. Kane, Phys. Rep. 117 (1985) 75.

\bibitem{R2} For a review on the Higgs sector of the SM and the MSSM ,
see J.F. Gunion, H.E. Haber, G.L. Kane and S. Dawson, {\it The Higgs 
Hunter's Guide}, Addison--Wesley, Reading 1990.

\bibitem{R3}
Y. Okada, M. Yamaguchi and T. Yanagida, Prog. Theor. Phys. 85 (1991) 1; \\
H. Haber and R. Hempfling, Phys. Rev. Lett. 66 (1991) 1815; \\
J. Ellis, G. Ridolfi and F. Zwirner, Phys. Lett. 257B (1991) 83; \\
R. Barbieri, F. Caravaglios and M. Frigeni, Phys. Lett. 258B (1991) 167.

\bibitem{R4} A. Djouadi, J. Kalinowski and P. Zerwas, Z. Phys. C70 (1996) 435.

\bibitem{R5}  V. Barger, M. Berger and P. Ohmann, Phys. Rev. D49 (1994)
4908; \\ W. de Boer, A. Dabelstein, W. Hollik, W. M\"osle and U. Schwickerath,
hep-ph/9609202.

\bibitem{R6a} 
H. Baer, D. Dicus, M. Drees and X. Tata, Phys. Rev. D36 (1987) 1363; \\
J. Gunion and H. Haber, Phys. Rev. D37 (1988) 2515; \\
A. Djouadi, J. Kalinowski and P. Zerwas, Z. Phys. C57 (1993) 569; \\
J. Gunion and J. Kelly, hep-ph/9610495. 

\bibitem{R6b} A. Djouadi, J. Kalinowski, P. Ohmann and P. Zerwas, 
hep-ph/9605339; \\ A. Bartl et al., hep-ph/9607388. 

\bibitem{R7} E. Brateen and J.P. Leveille, Phys. Rev. D22 (1980) 715; \\
M. Drees and K. Hikasa, Phys. Lett. B240 (1990) 455; \\
A. Mendez and A. Pomarol, Phys. Lett. B252 (1990) 461; \\
A. Djouadi and P. Gambino, Phys. Rev. D51 (1995) 218; \\ 
A. Djouadi, M. Spira and P.M. Zerwas, Phys. Lett. B264 (1991) 440; \\
M. Spira et al., Nucl. Phys. B453 (1995) 17.

\bibitem{R8} A. Bartl et al., Phys. Lett. B378 (1996) 167. 

\bibitem{R10a}  A. Bartl, W. Majerotto and W. Porod, Z. Phys. C64 (1994) 499. 

\bibitem{R9} Particle Data Group, Phys. Rev. D54 (1996) 1. 

\bibitem{R10} J. Ellis and D. Rudaz, Phys. Lett. B128 (1983) 248.

\bibitem{R11} J. Gunion and H. Haber, Nucl. Phys.  B307 (1988) 445.

\bibitem{R12} W. Siegel, Phys. Lett. B84 (1979) 193; \\
D. M. Capper, D.R.T. Jones, P. van Nieuwenhuizen, Nucl. Phys. B167 (1980)
479. 

\bibitem{R13} G. 't Hooft and M. Veltman, Nucl. Phys. B44 (1972) 189;\\
P. Breitenlohner and D. Maison, Commun. Math. Phys. 52 (1977) 11.

\bibitem{R14} S. Martin and M. Vaughn, Phys. Lett. B318 (1993) 331.
   
\bibitem{R15} G. Passarino and M. Veltman, Nucl. Phys. B160 (1979) 151;
\\  G. 't Hooft and M. Veltman, Nucl. Phys. B153 (1979) 365;\\
G.J. Oldenborgh, Comput. Phys. Comm. 66 (1991) 1. 

\bibitem{R16a}
A. Djouadi, W. Hollik and C. J\"unger, Phys. Rev. D54 (1996) 5629; \\
A. Djouadi, W. Hollik and C. J\"unger, hep-ph/9609419, Phys. Rev. D to appear.

\bibitem{R16b} A. Bartl, H. Eberl, W. Majerotto, Nucl. Phys. B472 (1996) 481; \\
W. Beenakker, R. H\"opker, T. Plehn and P.M. Zerwas, hep-ph/9610313.

\bibitem{R17} K. Hikasa and Y. Nakamura, Z. Phys. C70 (1996) 139.

\bibitem{den} A. Denner, Fortschr. Phys. 41 (1993) 4.

\bibitem{R18} A. Bartl et al., hep-ph/9701398.

\end{thebibliography}
\end{document}